\documentclass[twocolumn,prl,aps]{revtex4}

\usepackage{graphicx}
\usepackage{amsmath}
\usepackage[french]{babel}
\usepackage[latin1]{inputenc}

\begin{document}

\title{Comment on the paper by Rovelli \& al. about the compatibility\\ of various  "gauge conditions"}
\author{Germain Rousseaux}
\affiliation{Universit\'e de Nice Sophia-Antipolis,\\
Institut Non-Lin\'{e}aire de Nice,\\
UMR 6618 CNRS-UNSA, \\ 
1361, route des Lucioles. 06560 Valbonne, France. \\
(Germain.Rousseaux@inln.cnrs.fr) }

\begin{abstract}
The compatibility "demonstrated" by Rovelli \& al. in \cite{Rovelli} between various "gauge conditions" both in Classical Electromagnetism and General Relativity can be better understood if one distinguishes "gauge conditions" of the solution type and "gauge conditions" of the constraint type.
\end{abstract}

\maketitle

Recently, we showed  that the well-known Coulomb and Lorenz "gauge conditions"  (see \cite{Okun} for an introduction to the various "gauge conditions" encountered in the literature) were, in fact, not equivalent because they must be interpreted as physical constraints that is electromagnetic continuity equations \cite{AFLB03}. In addition, we were able to demonstrate that the Coulomb "gauge condition" is the galilean approximation of the Lorenz "gauge condition" within the magnetic limit of L\'evy-Leblond \& Le Bellac \cite{LBLL,EPL05,EJP06,AJP07}. So, to "make a gauge choice" that is choosing a gauge condition is, as a consequence of our findings, not related to the fact of fixing a special couple of potentials. Gauge conditions are completely uncorrelated to the supposed indeterminacy of the potentials. Hence, we propose to rename the "gauge conditions" like the ones of Lorenz or Coulomb by physical "constraints".

The purpose of our comment is to show that a "demonstration" of the compatibility between the Fock-Schwinger "gauge condition" and the Lorenz "gauge conditions" by Rovelli \& al. in \cite{Rovelli} is blurred by a slip in terminology. As a matter of fact, the Lorenz equation $\partial _\mu A^\mu =0$ is according to us a constraint whereas the Fock-Schwinger equation $x_\mu A^\mu =0$ is what we called a "solution" submitted to the Lorenz constraint. If so, the "compatibility" is obvious...

First of all, let us illustrate what we mean by a "solution" under a physical constraint. One often finds in textbooks that we can describe a uniform magnetic field ${\bf B} =B {\bf e}_{z}$ by either the so-called
symmetric "gauge" $\bf{A_1}=1/2\bf{B} \times \bf{r}$ or by the
so-called Landau "gauge" \cite{Okun}. This two "gauges" are related
by a gauge transformation :

\begin{equation}
{\bf A}_{1}=\frac{1}{2} {\bf B} \times {\bf r} = \frac{1}{2} [-By,Bx,0]
\end{equation}
becomes either :
\begin{equation}
{\bf A}_{2} =[0,Bx,0]
\quad
or
\quad
{\bf A}_{3}=[-By,0,0]
\end{equation}
with the gauge functions $\pm f = \pm xy/2$.

However, there is no discussion in the litterature of the following 
issue. As a matter of fact, if we consider a solenoid with a current along
$\bf{e_\theta}$, the magnetic field is uniform (along $\bf{e_z}$)
and could be described by the symmetric "gauge" or the Landau
"gauge". Yet, the vector potential in the Landau "gauge"  ${\bf A}_{2}$ is along
$\bf{e_y}$ whereas the vector potential in the symmetric gauge is
along $\bf{e_\theta}$. We advocate that only the symmetric "gauge"
is valid in this case because it does respect the symmetry of the 
currents (${\bf J} = J {\bf e}_{\theta}$) whereas the Landau "gauge" 
does not. Moreover, the symmetric "gauge" (or the Landau "gauge") is not, in fact, a gauge condition
 but a solution describing a uniform magnetic field under the Coulomb constraint
($\nabla .{\bf{A_1}}=\bf{0}$).
In order to understand this last point, one can picture an analogy between Fluid Mechanics and Classical Electromagnetism. Indeed, the solenoid is analogous to a  cylindrical vortex core with vorticity $\bf{w}$ and we know that the velocity inside the core is given by
$\bf{u}=1/2\bf{w} \times \bf{r}$ which is analogous to the
symmetric gauge for an incompressible flow ($\nabla .{\bf{u}}=\bf{0}$). Outside the vortex core, the velocity is given by \cite{GHP} :
\begin{equation}
{\bf u}=\frac {\Gamma \nabla \theta} {2
\pi}=\frac {\Gamma}{2 \pi r} {\bf e_\theta}
\end{equation}
where $\Gamma$ is the flux of vorticity inside the vortex or the
circulation of the velocity outside the vortex. One recovers the
analogue formula for the vector potential outside a solenoid...

Of course, if the problem we are considering does not feature the cylindrical geometry (two horizontal plates with opposite surface currents for example, analogous to a plane Couette flow \cite{GHP}), one of the Landau "gauges" ${\bf A}_{2}$ or ${\bf A}_{3}$ submitted to the constraints $\nabla .{\bf{A_2}}=\bf{0}$ or $\nabla .{\bf{A_3}}=\bf{0}$ must be used instead of the symmetric "gauge" ${\bf A}_{1}$ according to the necessity of respecting the underlying distribution/symmetry of the currents which is at the origin of both the vector potential and the magnetic field. To give a magnetic vector field without specifying its current source is an ill-posed problem which was interpreted so far by attributing an indeterminacy to the vector potential which is wrong. Now, how can we test experimentally this argument based on symmetry ? If the current of the solenoid varies with time, it will create an electric field which is along $\bf{e_\theta}$ as the vector potential because the electric field is minus the time derivative of the vector potential. If the currents in the horizontal plates change with time, a horizontal electric field will appear for the same reason.

We come back now to the main claim of Rovelli \& al. \cite{Rovelli}. We will argue that the Fock-Schwinger equation is a  solution submitted to the Lorenz constraint. Let us write the Fock-Schwinger "gauge" $x_\mu A^\mu =0$ in projection for cartesian coordinates $x_0 A^0 +x_1 A^1+x_2 A^2 +x_3 A^3 =0$. The four-position and four-potential have components $x_\mu =\left (-ct,{\bf x} \right )$ and $A^\mu=\left (V/c,{\bf A} \right )$. For example, the Fock-Schwinger equation becomes $xA_x +yA_y +zA_z =tV$ in cartesian coordinates. 

The Lorenz "gauge" $\nabla .{\bf{A}}+1/c^2 \partial V/\partial t =0$ becomes after a Fourier transformation $i{\bf k . A}-i\omega /c^2 V=0$. But a light wave is such that $|k|=\omega /c$.

A plane or spherical light wave has a null space-time interval ($ds^2 =dx_\mu .dx^\mu =0$): $x^2 +y^2 +z^2 =c^2 t^2$ or $r^2=c^2 t^2$. The Lorenz constraint for a plane or spherical light wave becomes: $V=cA_x$ or $V=cA_r$ where $A_x$ or $A_r$ is the component of the light wave which is parallel to the wave-vector ${\bf k}$. With $x=ct$ ($y=z=0$) or $r=ct$ for outgoing waves, one gets $V=cA_x = x/t A_x$ or $V=cA_r =r/t A_r$ that is the Fock-Schwinger solution: $xA_x =tV$ or $rA_r =tV$.

As a conclusion, the Fock-Schwinger solution describes the propagation of a light wave submitted to the Lorentz-covariant Lorenz constraint which is the continuity equation allowing the propagation. It is interesting to notice that V. A. Fock himself derived the Fock-Schwinger equation starting with the Lorenz equation and using a gauge transformation \cite{Fock}...

\end{document}